\documentclass[tighten, times, twocolumn, trackchanges]{aastex62}

\usepackage{ifthen}
\usepackage{natbib}
\usepackage{amssymb, amsmath}
\usepackage{appendix}
\usepackage{etoolbox}
\usepackage[T1]{fontenc}
\usepackage{paralist}
\usepackage[normalem]{ulem}
\usepackage{ragged2e}
\usepackage{newtxtext}  
\usepackage{chemformula}
\usepackage{listings}
\usepackage{xcolor}
\usepackage[bottom]{footmisc}
\usepackage[flushleft]{threeparttable}

\hypersetup{citecolor=blue, 
            linkcolor=blue, 
            menucolor=blue, 
            urlcolor=blue}  

\providecommand{\adsurl}[1]{\href{#1}{ADS}}

\makeatletter
\patchcmd{\NAT@citex}
  {\@citea\NAT@hyper@{%
     \NAT@nmfmt{\NAT@nm}%
     \hyper@natlinkbreak{\NAT@aysep\NAT@spacechar}{\@citeb\@extra@b@citeb}%
     \NAT@date}}
  {\@citea\NAT@nmfmt{\NAT@nm}%
   \NAT@aysep\NAT@spacechar\NAT@hyper@{\NAT@date}}{}{}

\patchcmd{\NAT@citex}
  {\@citea\NAT@hyper@{%
     \NAT@nmfmt{\NAT@nm}%
     \hyper@natlinkbreak{\NAT@spacechar\NAT@@open\if*#1*\else#1\NAT@spacechar\fi}%
       {\@citeb\@extra@b@citeb}%
     \NAT@date}}
  {\@citea\NAT@nmfmt{\NAT@nm}%
   \NAT@spacechar\NAT@@open\if*#1*\else#1\NAT@spacechar\fi\NAT@hyper@{\NAT@date}}
  {}{}
\makeatother

\DeclareSymbolFont{UPM}{U}{eur}{m}{n}
\DeclareMathSymbol{\umu}{0}{UPM}{"16}
\let\oldumu=\umu
\renewcommand\umu{\ifmmode\oldumu\else$\oldumu$\fi}
\newcommand\micro{\umu}
\renewcommand\micron{\micro m}
\newcommand\microns{\micron}

\newcommand\JWST{{\em JWST}}

\newcommand{\tso}{\textsc{Gen TSO}}
\newcommand{\pandeia}{\textsc{Pandeia}}
\newcommand\pandexo{\textsc{PandExo}}
\newcommand{\synphot}{\textsc{synphot}}
\newcommand\pyratbay{\textsc{Pyrat Bay}}

\newenvironment{myitemize}
{\begin{itemize}
    \setlength{\itemsep}{1pt}
    \setlength{\parskip}{0pt}
    \setlength{\parsep}{0pt}
}
{\end{itemize}}

\definecolor{codegreen}{rgb}{0,0.6,0}
\definecolor{codegray}{rgb}{0.5,0.5,0.5}
\definecolor{codepurple}{rgb}{0.58,0,0.82}
\definecolor{backcolour}{rgb}{0.95,0.95,0.92}

\lstdefinestyle{mystyle}{
    backgroundcolor=\color{backcolour},
    commentstyle=\color{codegreen},
    keywordstyle=\color{magenta},
    numberstyle=\tiny\color{codegray},
    stringstyle=\color{codepurple},
    basicstyle=\ttfamily\footnotesize,
    breakatwhitespace=false,
    breaklines=true,
    captionpos=b,
    keepspaces=true,
    numbers=left,
    numbersep=5pt,
    showspaces=false,
    showstringspaces=false,
    showtabs=false,
    tabsize=2
}

\shorttitle{The Gen TSO exoplanet simulator}
\shortauthors{Patricio E. Cubillos}

\begin{document}

\title{{\tso}: A General JWST Simulator for Exoplanet Times-series Observations}

\author[0000-0002-1347-2600]{Patricio E. Cubillos}
\affiliation{Space Research Institute, Austrian Academy of Sciences,
    Schmiedlstrasse 6, A-8042, Graz, Austria}
\affiliation{INAF -- Osservatorio Astrofisico di Torino,
    Via Osservatorio 20, 10025 Pino Torinese, Italy}

\email{patricio.cubillos@oeaw.ac.at}

\begin{abstract}

{\tso} is a noise calculator specifically tailored to simulate {\em
James Webb Space Telescope} ({\JWST}) time-series observations of
exoplanets. {\tso} enables the estimation of signal-to-noise ratios
(S/N) for transit or eclipse depths through an interactive graphical
interface, similar to the {\JWST} Exposure Time Calculator (ETC).
This interface leverages the ETC by combining its noise simulator,
{\pandeia}, with additional exoplanet resources from the NASA
Exoplanet Archive, the Gaia DR3 catalog of stellar sources, and the
TrExoLiSTS database of {\JWST} programs. The initial release of {\tso}
allows users to calculate S/Ns for all {\JWST} instruments for the
spectroscopic time-series modes available as of the Cycle 4 GO
call. Additionally, {\tso} allows users to simulate target acquisition
on the science targets or, when needed, on nearby stellar targets
within the visit splitting distance. This article presents an overview
of {\tso} and its main functionalities. {\tso} has been designed to
provide both an intuitive graphical interface and a modular API to
access the resources mentioned above, facilitating planing and
simulation of {\JWST} exoplanet time-series observations. {\tso} is
available for installation via the Python Package Index and its
documentation can be found
at \href{https://pcubillos.github.io/gen\_tso}{pcubillos.github.io/gen\_tso}.

\end{abstract}

\section{Introduction}

Time-series observations, particularly transit and eclipse
spectroscopy, have become a cornerstone of exoplanet characterization.
The start of operations of the {\em James Webb Space Telescope}
(\JWST), has demonstrated this over the last couple of years.
Thanks to its unrivaled data quality and spectral coverage, {\JWST}
have revolutionized our ability to detect atmospheric molecules,
constrain cloud properties, and infer the dynamic, chemical, and
physical processes responsible for these observations \citep[e.g.,][]{
AhrerEtal2023naturJWSTersWASP39bNIRcam,
AldersonEtal2023naturJWSTersWASP39bG395H,
CoulombeEtal2023naturWASP18bEmissionNIRISS,
BeattyEtal2024apjGJ3470bSO2, BellEtal2024natasWASP43bMIRIphase,
BellEtal2023naturWASP80bMethaneNIRCam,
BennekeEtal2024arxivTOI270dMiscible,
CarterEtal2024natasDataSynthesisWASP39b,
DyrekEtal2024natWASP107bTransitMIRI,
FeinsteinEtal2023naturJWSTersWASP39bNIRISS,
FuEtal2024natasHD189733bH2S,
HolmbergMadhusudhanEtal2024aaTOI270dJWSThycean,
MadhusudhanEtal2023apjK2-18bJWSTcarbonMolecules,
MurphyEtal2024natasLimbAsymmetryWASP107b,
RustamkulovEtal2023naturJWSTersWASP39bPRISM,
SingEtal2024natWASP107bJWSTnirspec,
ValentineEtal2024ajWASP17bJWSTmiri,
WelbanksEtal2024natWASP107bJWSTtransits,
XueEtal2024apjHD209458bJWSTtransits}.

To plan observations, the Space Telescope Science Institute provides
the {\JWST} Exposure Time Calculator
(ETC, \href{https://jwst.etc.stsci.edu/}{https://jwst.etc.stsci.edu/}),
a web-based tool for estimating the exposure time needed to achieve a
desired science goal.
The ETC uses the current best knowledge of the performance of the
{\JWST} instruments to estimate their flux rates, noise properties,
and signal-to-noise ratios.
As such, the ETC is a vital tool for simulating realistic exoplanet
time-series observations with {\JWST}.

However, the ETC is a complex tool, since it must support all {\JWST}
instruments and observing modes.  {\JWST} is equipped with four
instruments to sample the near-infrared (NIRISS, NIRCam, and NIRSpec,
0.6--5.0~{\microns}) and mid-infrared region of the spectrum (MIRI,
5.0--28.0~{\microns}), each with their own specific observing modes
including imaging, spectroscopy, coronography, and aperture masking
interferometry.
These instruments were designed to study a variety of astronomical
phenomena including the early universe after the Big Bang, the
evolution of galaxies, the formation of stars and their planetary
systems, and the physical properties of solar and extrasolar planets.

Further, estimating the signal-to-noise ratio (S/N) of exoplanet
time-series observations requires additional calculations beyond the
core design of the ETC, as well as additional information of stellar
and planetary parameters.  This information is distributed in multiple
databases, e.g., the NASA Exoplanet
Archive\footnote{\href{https://exoplanetarchive.ipac.caltech.edu}
{https://exoplanetarchive.ipac.caltech.edu}} for the planetary
physical parameters; the SIMBAD
database\footnote{\href{https://simbad.u-strasbg.fr/simbad}
{https://simbad.u-strasbg.fr/simbad}}, Gaia DR3
catalog\footnote{\href{https://gea.esac.esa.int/archive/}
{https://gea.esac.esa.int/archive}}, and 2MASS
catalog\footnote{\href{https://irsa.ipac.caltech.edu/Missions/2mass.html}
{https://irsa.ipac.caltech.edu/Missions/2mass.html}} for the for
stellar physical properties and coordinates; or the TrExoLiSTS
database\footnote{\href{https://www.stsci.edu/~nnikolov/TrExoLiSTS/JWST/trexolists.html}
{https://www.stsci.edu/$\sim$nnikolov/TrExoLiSTS/JWST/trexolists.html}} for
{\JWST} observing programs.
Using the ETC for exoplanet applications can thus be complex and time
consuming.

\begin{figure*}[t]
\centering
\includegraphics[width=0.97\linewidth]{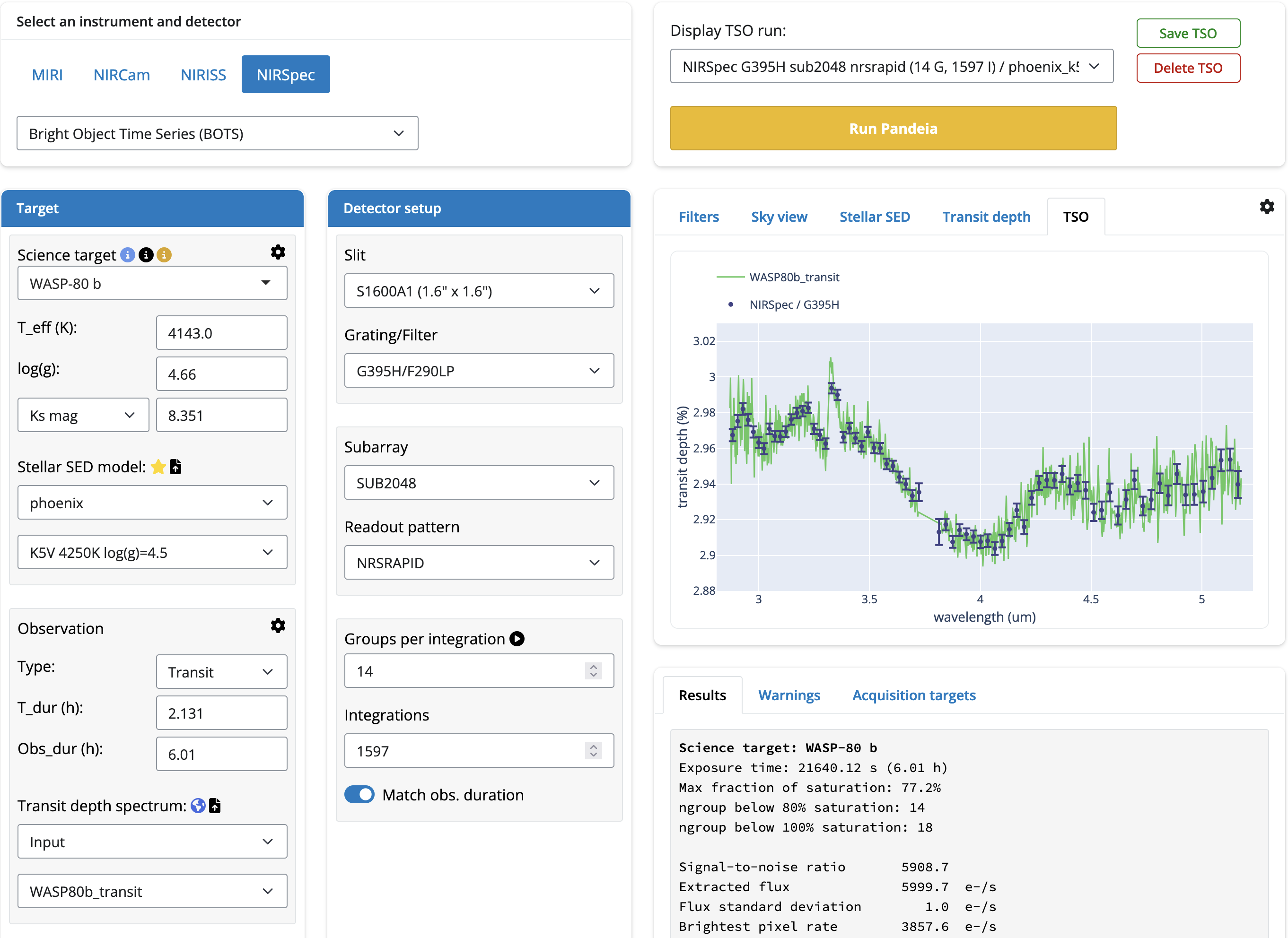}
  \hspace{2cm}
\caption{
  {\tso} graphical user interface.  This is a layout that incorporates
  thes elements of the {\JWST} ETC, but custom designed for transiting
  exoplanet time-series observations.  The layout has well defined
  sections for inputs and outputs, as well as astrophysical and
  instrumental settings.
  The top left panel sets the instrument and mode selection.
  The lower left panel (``Target'') contains the inputs
  for the target, host star, and transit or eclipse settings.
  The panel to the right (``Detector setup'')  contains the inputs
  for the instrumental configuration.  The top panel on the right 
  contains the buttons to calculate S/N
  estimations and list of all previous runs.  Below there are two
  multi-tab panels with ouput figures (filters, field of view, stellar SED
  models, transit or eclipse spectra, and time-series S/N
  simulations) and output text information (real-time S/N, warnings, and list of
  stellar sources around the science target).}
\label{fig:gen_tso_gui}
\end{figure*}

With this in mind, we have developed {\tso}, an open-source
Python package designed to simplify the estimation of exoplanet
transit and eclipse S/Ns from {\JWST} time-series observations.
Gen TSO provides a streamlined graphical interface that emulates the ETC web
application, but containing only the components
that are needed for exoplanet simulations.  This interface is
enhanced by incorporating physical and observational information for
the known population of exoplanets from a variety or databases,
enabling the direct calculation of transit or eclipse depths S/Ns, and
providing a range of interactive visualization tools.

This article introduces {\tso} package to the wider scientific
community.  Section \ref{sec:overview} outlines the components,
features, and usage of {\tso}.  Section \ref{sec:validation}
provides benchmark comparisons to validate the package's
calculations.  Finally, Section \ref{sec:conclusions} presents a summary and future development plans.

\section{Overview of Gen TSO}
\label{sec:overview}

At its core, the {\tso} package is an exoplanet noise simulator for
{\JWST}. Its main interface is a graphical user application (Figure
\ref{fig:gen_tso_gui}). This interface resembles the {\JWST} ETC;
However, it is simplified by focusing specifically on the simulation
of time-series observations. Importantly, {\tso} also leverages the
ETC by offering additional features, such as direct access to several
exoplanet databases and the ability to directly estimate S/Ns for
exoplanet transits and eclipses. This is achieved by bringing together
{\pandeia} \citep{PontoppidanEtal2016spiePandeia}, the ETC's noise
calculator, with the following resources:

\begin{myitemize}
    \item {\synphot} \citep{STScI2018asclSynphot}: stellar spectral
    energy distribution (SED) models
    
    \item NASA Exoplanet Archive: system parameters of confirmed and
    candidate exoplanets

    \item TrExoLiSTS \citep{NikolovEtal2022rnaasTrExoLiSTS}: list of
    approved {\JWST} exoplanet time-series programs

    \item Gaia Data Release
    3 \citep[DR3,][]{GaiaCollab2023aaGaiaDR3Summary}: catalog of
    stellar sources

    \item ESASky \citep{BainesEtal2017paspESASky,
    GiordanoEtal2018acESASky}: interactive visualization of the field
    of view
\end{myitemize}

The goal of the {\tso} package is to provide an intuitive and
interactive interface, integrating in a single place the resources
required to simulate {\JWST} observations. The following sections
describe in more detail the graphical interface and the features
enabled by each of these components.

\subsection{Graphical User Interface}
\label{sec:gui}

The {\tso} graphical user interface (Figure \ref{fig:gen_tso_gui}) is a
Python-based application deployed locally in the browser using
the \textsc{Shiny} package.  \textsc{Shiny} provides a highly reactive
framework that enables an interactive interface, allowing users to
visualize inputs and outputs in real-time, upload and download files,
and perform web-based operations.

The graphical interface is divided into three main sections for the
physical inputs, instrumental inputs, and outputs.
At the top of the interface, the user can select the instrument and
observing mode.  {\tso} simplifies this choice by only including the
{\JWST} observing modes that can perform spectroscopic time-series
observations and target acquisition.  Table \ref{table:gen_tso_modes}
list the available observing modes for each instrument at the time of
publication.

\begin{table}[h]
\centering
\caption{Available observing configurations in {\tso}}
\label{table:gen_tso_modes}
\begin{tabular*}{1.0\linewidth} {@{\extracolsep{\fill}} ll}
\hline
\hline
Instrument & Modes \\
\hline
{\bf NIRISS} & Single Object Slitless Spectroscopy (SOSS) \\
   & Target Acquisition \\
\hline
{\bf NIRSpec} & Bright Object Time Series (BOTS) \\
   & Target Acquisition \\
\hline
{\bf NIRCam}  &  LW Grism Time Series   \\
   & SW Grism Time Series   \\
   & Target Acquisition \\
\hline
{\bf MIRI}   &  Low Resolution Spectroscopy (LRS) Slitless \\
   & MRS Time Series \\
   & Target Acquisition \\
\hline
\hline
\end{tabular*}
\end{table}

As the user selects an observing mode, the ``Detector setup'' panel
(Fig.\ \ref{fig:gen_tso_gui}) updates the instrumental configuration
options for the current mode. These include the choice of disperser,
filter, orders (for NIRISS), subarray, readout pattern, groups, and
integrations.

\begin{figure}[t]
\includegraphics[width=\linewidth,clip]{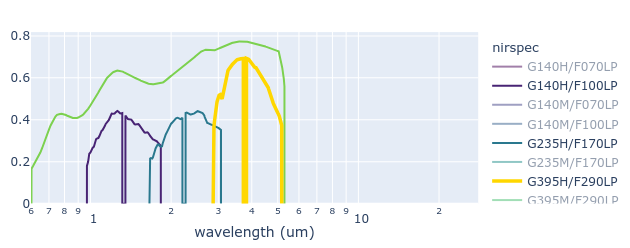}
\includegraphics[width=\linewidth,clip]{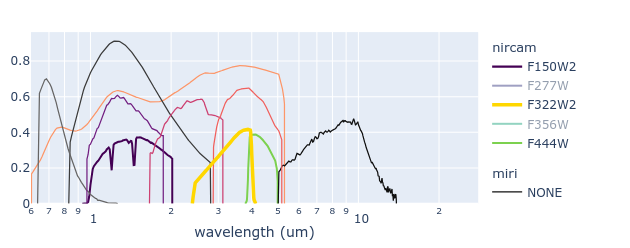}
\caption{
  Filter throughput display for the NIRSpec instrument (top) or
  for all instruments (bottom).  This display highlights the
  throughput currently selected (yellow thicker line) in addition to
  other available throughputs, helping the user to visualize the
  spectral range being covered and compare to that of other
  configurations.  Users can interactively select/unselect which
  filters to display with the menu on the right hand side of the
  display.}
\label{fig:gen_tso_filters}
\end{figure}

To help users to quickly assess the spectral range covered by the
combination of disperser, filter, and subarray, the panel on the right
(``Filters'' tab) interactively displays the throughput curves of
different instruments and observing modes, highlighting the 
throughput for the current configuration (Fig.\
\ref{fig:gen_tso_filters}).

The ``Target'' panel on the left of the interface (Fig.\
\ref{fig:gen_tso_gui}) contains the
configuration for the stellar and planetary properties.  At the top of
the panel a menu allows users to search for known science targets.
Selecting one fills automatically the rest of the fields in this panel
with information extracted from the NASA Exoplanet Archive.
This menu
contains all known confirmed and TESS-candidate targets. If desired,
users can filter the target selection to transiting, non-transiting,
candidates, or previously observed JWST targets.
Section \ref{sec:exo_catalog} explains in more detail how this
information is extracted and maintained.

The buttons above the selected target provide additional information
about the selected target, such as name aliases, a link to the NASA
archive, a link to the system parameters, and a link to the previous
JWST programs of the host star.

\begin{figure*}[t]
\centering
\includegraphics[width=0.92\linewidth,clip]{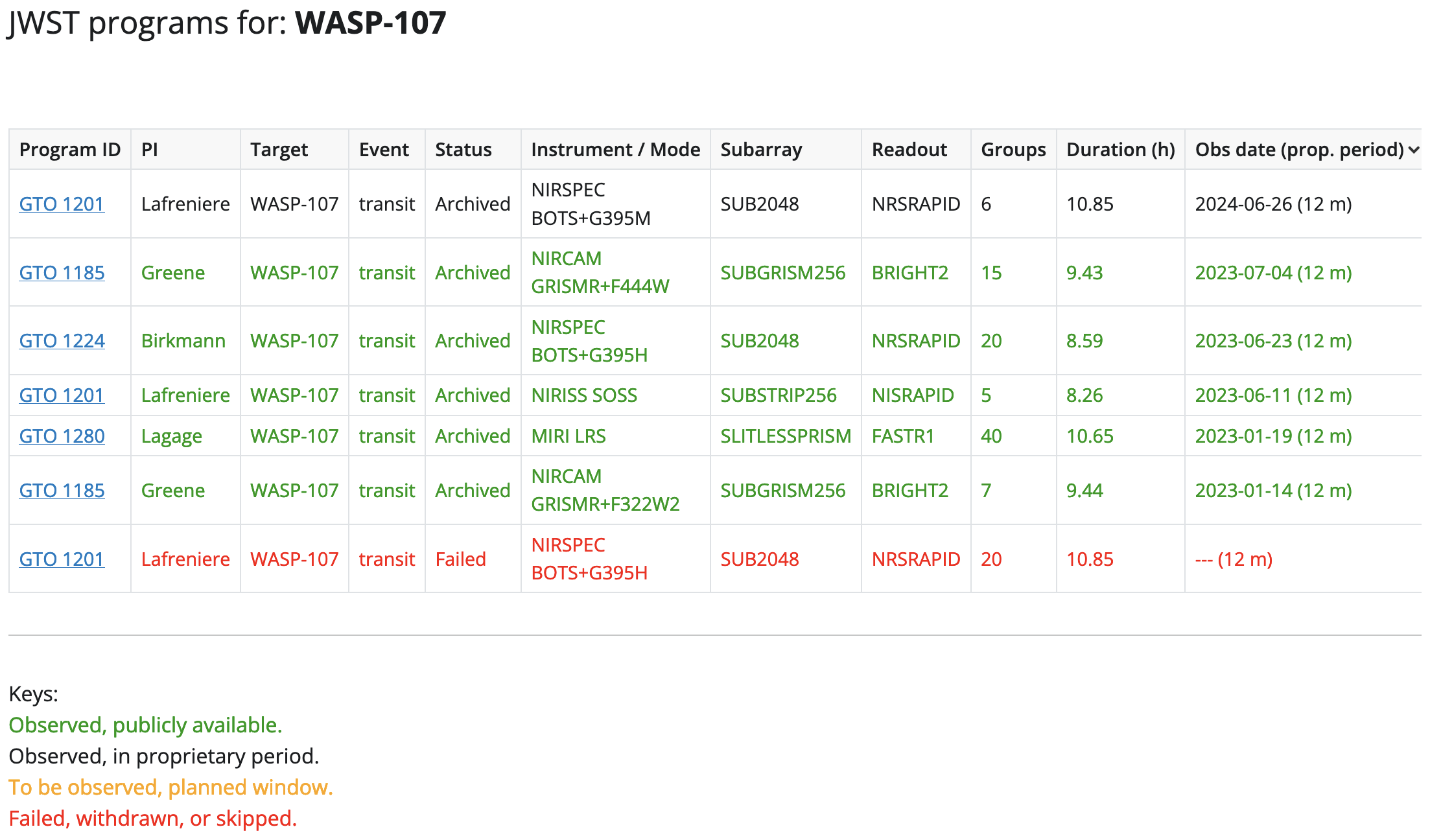}
\caption{
  List of {\JWST} observing programs for the host star WASP-107.  For
  each observation, {\tso} displays the program ID and principal investigator, type of
  observation, instrumental configuration, and observing timing.  To
  guide the users, the observations are color coded by publicly
  available (green), in proprietary period (black), to be observed
  (orange), and unsuccessful (red). This information helps users to
  avoid duplicating observations, complement observation with other
  instruments, or simulate existing observations. The Program ID
  column provides links to the programs, hosted at the Space Telescope
  website.}
\label{fig:gen_tso_programs}
\end{figure*}

\subsection{NASA Exoplanet Archive}
\label{sec:exo_catalog}

{\tso} provides system
parameters for all confirmed exoplanets and TESS candidates, sourced
from the NASA Exoplanet Archive. Ideally, this information could be
retrieved live from the NASA web interface. However, the latency of
NASA online service makes this sub-optimal for a real-time interactive
application, as individual requests require several seconds to
execute. Therefore, we opted to fetch and integrate the NASA data
directly into the {\tso} package.

The advantage of this approach is that the application can switch
between targets in a seemingly instantaneous manner, as well as
function offline. The disadvantage is that not all targets may have
up-to-date parameters. To address this issue, the {\tso} package will
be updated regularly to include the latest system
parameters (approximately a monthly base, but more
frequently when expecting higher demand from proposal cycles).
Additionally, {\tso} provides the tools for users to fetch the current
values from the NASA database at any time.

To fetch the NASA data, we use their Table Access Protocol
service\footnote{\href{https://exoplanetarchive.ipac.caltech.edu/docs/TAP/usingTAP.html}
{https://exoplanetarchive.ipac.caltech.edu/docs/TAP/usingTAP.html}}
(TAP). TAP allows users to download the entire Planetary Systems and
TESS Project Candidates tables. From these we retrieve the host star
and planet names; the stellar radius, mass, Ks magnitude, effective
temperature, surface gravity, right ascension, and declination; and
the planetary radius, mass, orbital period, semi-major axis, and
transit duration. TESS candidates flagged as ``false alarm'' or
``false positive'' are excluded. Since the TESS table does not include
the Ks magnitudes, we obtained the values from the SIMBAD astronomical
database or the Vizier/2MASS catalog
\citep{WengerEtal2000aaSimbad}.

When multiple references (entries) exist for a given target, the
system parameters are selected adopting a hierarchical approach.
First, the parameters from the entry flagged as the ``default
parameter set'' by the NASA catalog are chosen.  If any
parameters are missing, they are supplemented with data from the
remaining entries, which have been sorted from most to least
complete (regarding the parameters listed in the paragraph
above).
Parameters of planets belonging to a same host are cross checked to
ensure consistent stellar parameters, adopting the values
from the most complete parameter set, or else from the one with most
entries (assuming that translates into the one most studied).  If
stellar parameters have changed, the planetary parameters are updated
while preserving the observed properties (e.g., if a stellar radius
changed, the planet radius is updated according to the $R_{\rm
p}/R_\star$ ratio).
The graphical interface provides a link to the
planet's entry in the NASA archive for users to verify the source
of the parameters shown (Fig.\ \ref{fig:gen_tso_gui}, black
circled {\em i} icon).

When dealing with entire catalogs, a main challenge is to ensure
consistency between different databases, primarily because objects are often
named differently across the multiple databases
(e.g., the same target is listed as WASP-178 in the NASA
confirmed-planets catalog, but is listed as TOI-1337 in the NASA
TESS-candidates catalog, and is listed as HD~134004 in the SIMBAD and
TrExoLiSTS databases; whereas in the literature the object can also be
found under the name KELT-26).
To solve this issue, we also collected all
known aliases from the NASA System Aliases Lookup Service, which we
complemented with the alias names from SIMBAD.  This allowed
us to identify and cross-match most targets, except for a few
non-standard names in TrExoLiSTS which were corrected manually.
As a result, targets searches in {\tso} will recognize any of the names
used in the NASA and TrExoLiSTS databases, as well as any of the
most common names found in the literature (i.e., HAT, KELT, MASCARA,
TOI, TrES, WASP, and XO).  For example, a search for
either ``WASP-189'' or ``HD 133112'' will return the target's {\JWST}
program information in {\tso}, even though that would not occur in a
search in TrExoLiSTS (at the time of publication).

\subsection{JWST exoplanet time-series programs}
\label{sec:trexo}

To provide the information on existing {\JWST} programs for a given
target, {\tso} integrates data from the TrExoLiSTS
database \citep{NikolovEtal2022rnaasTrExoLiSTS}.  This database
catalogs all accepted {\JWST} exoplanet time-series programs,
including executed and upcoming observations.  Each entry details the
target host, event type (transit, eclipse, or phase curve),
observation status (archived, pending, or unsuccessful), instrumental
configuration, observing date, program ID, and principal investigator.
{\tso} provides this data for each host, indicating which observations
are scheduled to be observed, under proprietary period, publicly
available, or were unsuccessful
(Fig.\ \ref{fig:gen_tso_programs}).

\subsection{Stellar SED models}

\begin{figure}[t]
\includegraphics[width=\linewidth,clip]{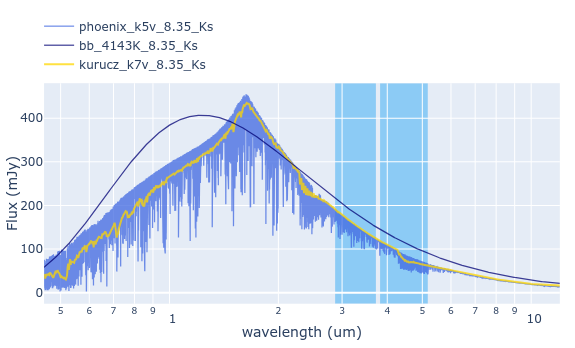}
\caption{
  Stellar SED viewer panel of the {\tso} graphical interface
  (Fig.\ \ref{fig:gen_tso_gui}).  The solid curves show a PHOENIX (K5V,
  light blue), Kurucz (K7V, yellow), and a blackbody ($T_{\rm eff}=4143$~K,
  yellow) SED models for the host star WASP-80.  All models are
  normalized to a Ks magnitude of 8.351.  The light-blue shaded area
  marks the spectral coverage of the filter currently selected in the
  application (NIRSpec G395H/F290P).}
\label{fig:gen_tso_sed}
\end{figure}

The spectral energy distribution (SED) of the host star is a key
factor determining the S/N of the observations. {\tso} enables four
types of SED models, similar to those available in the ETC.
These options are PHOENIX \citep{AllardHauschildt1995apjPHOENIX} and
Kurucz (1993) stellar models
\citep{Kurucz1979apjsStellarAtmosphereModels} both provided via the
{\synphot} package \citep{STScI2018asclSynphot}, blackbody spectra,
and custom-provided SED models. The later can be either pre-loaded or
uploaded while running the application. While selecting SED
models, users can bookmark them to visualize and compare in the
``Stellar SED'' tab (Fig.\ \ref{fig:gen_tso_sed}), facilitating an
easy comparison of different models.

When a science target is selected, the SED model is automatically set
to the most suitable option, depending on the SED type. PHOENIX and
Kurucz SEDs are updated to the one that best matches the target's
effective temperature ($T_{\rm eff}$) and surface gravity ($\log(g)$).
Blackbody SEDs adopt the effective temperature of the \texttt{T\_eff}
field.

\subsection{{\pandeia} Noise Modeling}

To compute the noise properties of an observation {\tso} uses the
{\pandeia} engine \citep{PontoppidanEtal2016spiePandeia}, the exposure
time calculator (ETC) system developed for the James Webb Space
Telescope (JWST). {\pandeia} calculates the two-dimensional
pixel-by-pixel signal and noise properties of the JWST instruments,
appropriately accounting for the instrumental point spread functions,
detector readouts, and detector readnoise.
The {\pandeia} photometric and spectral extraction strategies provide
the flux rates ($F(\lambda)$, in $e^-$\,s$^{-1}$) and their variances.

For exoplanet time-series observations, we are often
interested in the transit or eclipse depth:
\begin{equation}
\label{eq:depth}
d(\lambda) = 1 - F_{\rm in}/F_{\rm out}.
\end{equation}

For transits, the flux rates in and out of transit are
given by
\begin{eqnarray}
F_{\rm in} &=& F_{\star}(1-(R_{\star}/R_{\rm p})^2), \\
F_{\rm out} &=& F_{\star}.
\end{eqnarray}

For eclipses, the flux rates are given by
\begin{eqnarray}
F_{\rm in} &=& F_{\star}, \\
F_{\rm out} &=& F_{\star}+F_{\rm p}.
\end{eqnarray}

Following the last-minus-first formalism, {\tso} estimates the
time-integrated variances in and out of transit as the sum in
quadrature of the shot, background, and read noise: $\sigma^2
= \sigma^2_{\rm shot} + \sigma^2_{\rm bkg} + \sigma^2_{\rm
read}$ \citep[for a detailed description
see][]{BatalhaEtal2017paspPandexo}.  All of these quantities are
directly provided by {\pandeia} or can be derived from its outputs;
thus, all that remains is to propagate uncertainties according to the
transit or eclipse depth calculation of Eq.\ (\ref{eq:depth}):

\begin{equation}
\sigma_d^2 = \left(\frac{1}{t_{\rm in}F_{\rm out}}\right)^2 \sigma_{\rm in}^2
+ \left(\frac{F_{\rm in}}{t_{\rm out}F_{\rm out}^2}\right)^2 \sigma_{\rm out}^2,
\end{equation}
where $t_{\rm in}$ and $t_{\rm out}$ are the times spent integrating
in and out of transit, i.e., the time between the first and
last measurements of an integration multiplied by the
number of integrations \citep{RauscherEtal2007paspJWSTdetectors}.
These times are also a direct output from {\pandeia}, and depend on
the instrumental configuration.

\subsection{Time-series S/N Calculations}

To run transit or eclipse S/N calculations, users must configure the
observation in the ``Target'' panel of the application (Fig.\
\ref{fig:gen_tso_gui}). The timing parameters to set are the transit
duration and the observation duration. The default value of the
observation is automatically set according to the recommendation of
the {\JWST} documentation\footnote{\href{https://jwst-docs.stsci.edu/jwst-near-infrared-imager-and-slitless-spectrograph/niriss-example-science-programs/niriss-soss-time-series-observations-of-a-transiting-exoplanet/step-by-step-etc-guide-for-niriss-soss-time-series-observations-of-a-transiting-exoplanet}{https://jwst-docs.stsci.edu/jwst-near-infrared-imager-and-slitless-spectrograph/niriss-example-science-programs/niriss-soss-time-series-observations-of-a-transiting-exoplanet/step-by-step-etc-guide-for-niriss-soss-time-series-observations-of-a-transiting-exoplanet}}:

\begin{equation}
T_{\rm dur} = T_{\rm start} + T_{\rm settling} + T_{\rm base} + T_{\rm transit} + T_{\rm base},
\end{equation}

where $T_{\rm start}$ (defaulted value of 1~h) is the required start
timing window, $T_{\rm settling}$ (0.75~h) is the time recommended for
systematics settling, $T_{\rm transit}$ is the transit or eclipse
duration, and $T_{\rm base}$ ($\max\{0.5T_{\rm transit}, 1~{\rm h}\}$) is the
baseline time recommended before and after the transit. These default
parameters can be adjusted by the user.

\begin{figure*}[t]
\centering
\includegraphics[width=0.46\linewidth,clip]{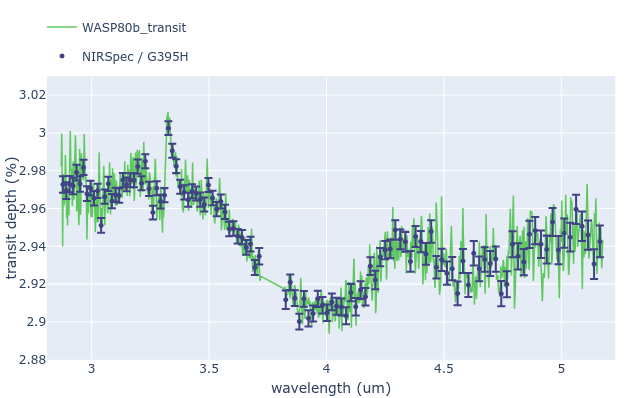}\hfill
\includegraphics[width=0.46\linewidth,clip]{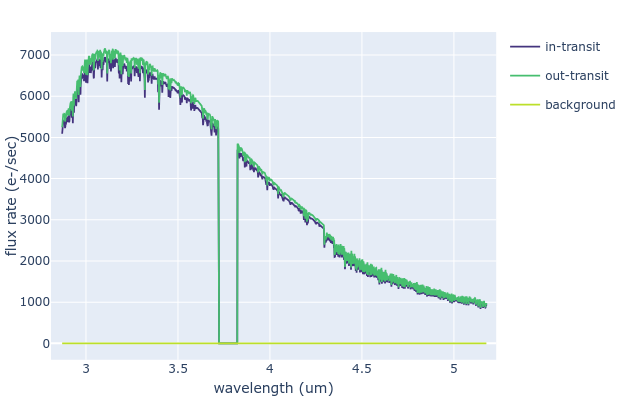}
\includegraphics[width=0.46\linewidth,clip]{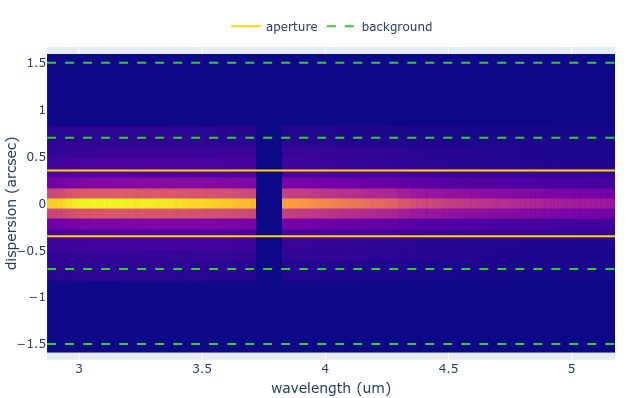}\hfill
\includegraphics[width=0.46\linewidth,clip]{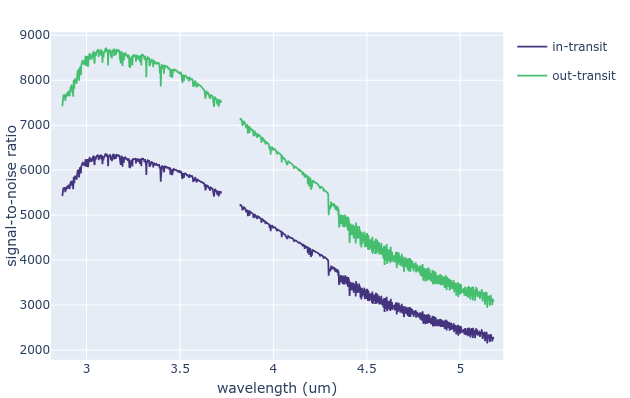}
\caption{
  Output plots for the simulation of a time-series observation of a
  WASP-80b transit with NIRSpec G395H/L290P. {\bf Top left:}
  Transmission spectrum of the {\JWST} simulation (blue markers with
  error bars) and the input model (solid green curve). {\bf Top
  right:} 1D flux rates of the source in-transit source,
  out-of-transit source, and background.  {\bf Bottom left:} 2D
  detector S/N map. The yellow and green curves denote the boundaries
  of the source and background used for the spectral extraction. {\bf
  Bottom right:} 1D S/N spectra of the source in and out of transit
  (note that these have been integrated in time, unlike the figure at
  the top).  }
\label{fig:gen_tso_pandeia_plots}
\end{figure*}

Models for the transit or eclipse depth spectra must be provided by
the users. These are plain text files containing the wavelength and
the depths in two columns. These files can be pre-loaded or updated
while running the application. Like with the SED models, the planetary
spectra can also be displayed and compared in the output panel.
Alternatively, {\tso} provides basic blackbody eclipse models or flat
transit spectra to make quick S/N assessments.

While setting the instrumental and astrophysical parameters, {\tso}
provides real-time feedback to set the number of groups and
integrations.  The ``Results'' tab (Fig.\ \ref{fig:gen_tso_gui}, lower
right panel) indicates the saturation level according to the current
number of groups, as well as the number of groups to reach 80\% and
100\% of saturation.  Clear warnings will be displayed when saturation
fractions exceed 100\%.
For a given SED, instrumental configuration, and exposure time, only
two scalar parameters are needed to characterize the saturation level:
the flux rate at the brightest pixel and the number of counts to
saturate the detector (computed by {\pandeia}).  {\tso} has tabulated
these values for the spectroscopic observing modes and for the PHOENIX
and Kurucz SEDs over a grid of Ks magnitudes (from 20 to 3.5),
allowing it to provide immediate information about the saturation
levels.  For other configurations, users can compute the saturation
information before deciding to calculate a full S/N estimation.
Additionally, the number of integrations can
be set to automatically match the total observation duration.

\begin{figure}[t]
\centering
\includegraphics[width=\linewidth,clip]{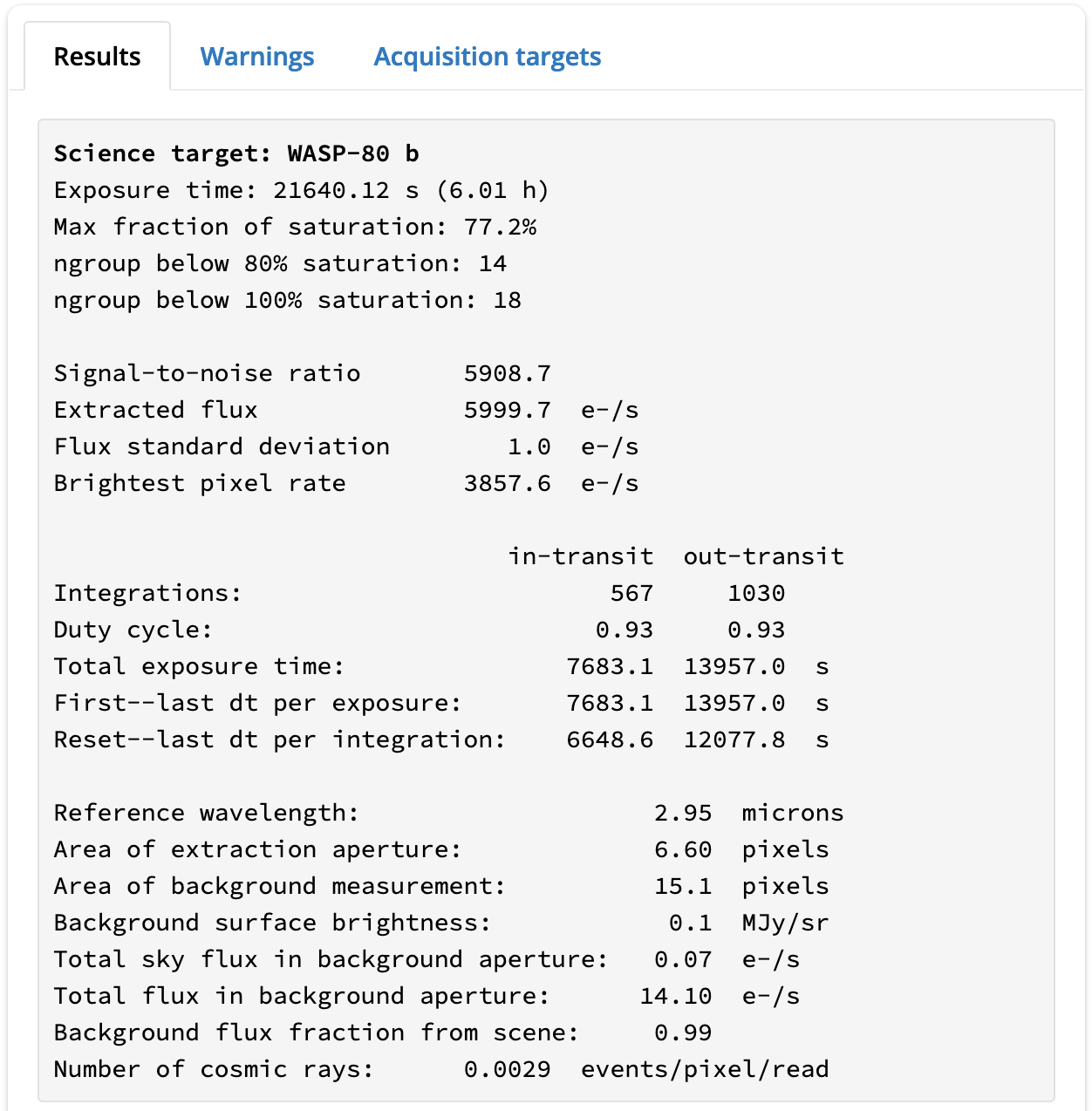}
\caption{
  Results output text for a WASP-80b transit simulation with NIRSpec
  G395H/L290P.}
\label{fig:gen_tso_text_results}
\end{figure}

After executing transit, eclipse, or acquisition S/N calculations
(Fig.\ \ref{fig:gen_tso_gui}, ``Run Pandeia'' button), the
application will store the configuration and outputs (which can then
be saved to a file for later use). The ``TSO'' tab will also display
the simulated observations, as well as
figure of the 1D and 2D flux rates, 1D and 2D S/N ratios, and 2D
saturation maps (Fig.\ \ref{fig:gen_tso_pandeia_plots}). All of these
figures are interactive, allowing users to change the data binning
resolution or number of simulated events.
Finally, a text output (similar to that of the ETC, Fig.\ \ref{fig:gen_tso_text_results}) will summarize the
simulation flux rates, S/Ns, integrations, duty cycle, exposure times,
and other values.  When applicable, this information will
be broken down by in/out-of-transit.

\subsection{Target Acquisition, ESASky viewer, and Gaia DR3 Catalog}
\label{sec:acquisition}

\begin{figure}[t]
\includegraphics[width=\linewidth,clip]{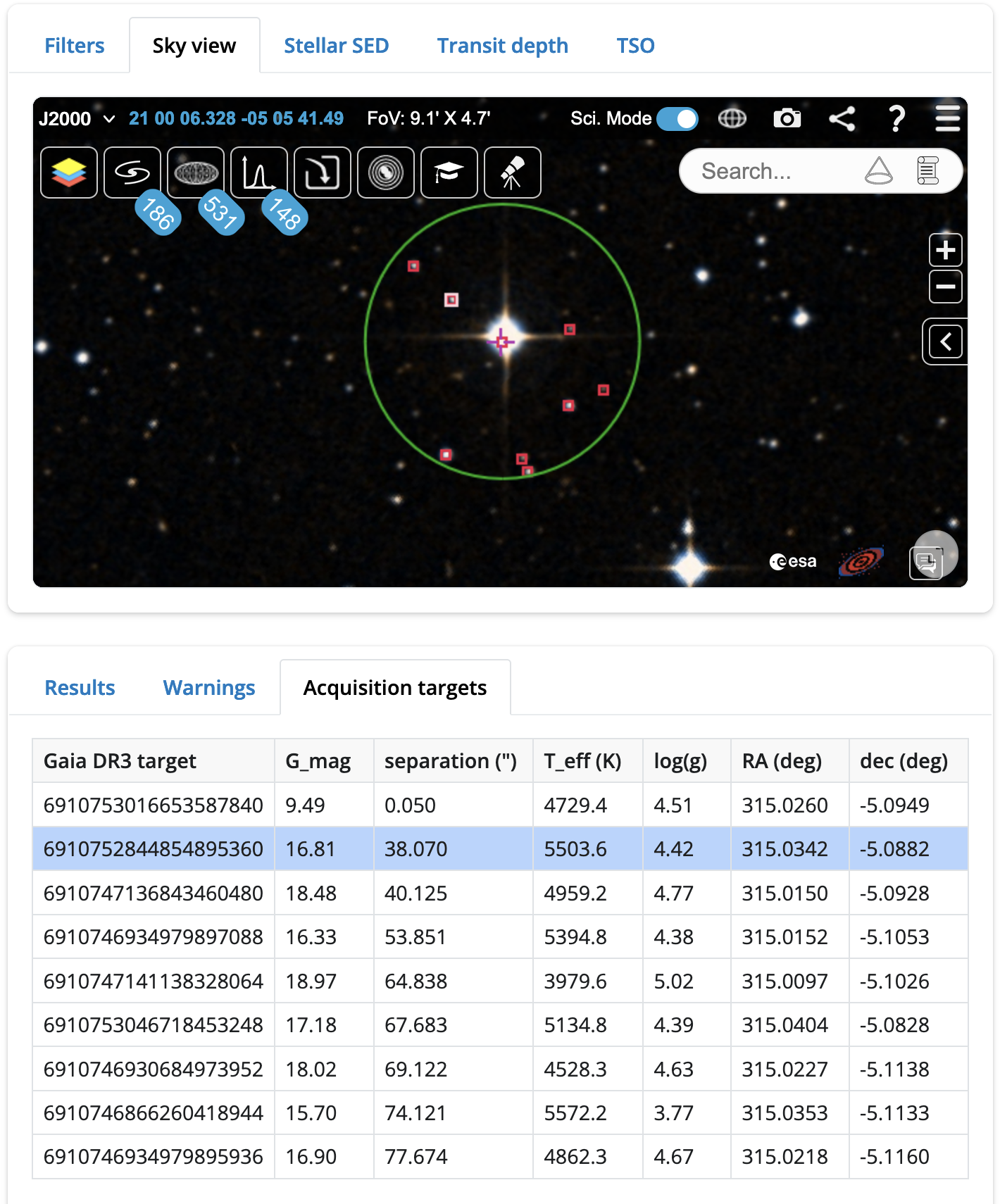}
\caption{
  {\bf Top:} ESASky visualization of the field of view around WASP-69.
  {\bf Bottom:} List of Gaia DR3 stellar sources within the visit
  splitting distance of WASP-69.  Queries for Gaia stellar sources
  around a target will trigger the red square markers in the ESASky
  viewer indicating the location of the Gaia stellar sources and the
  green circle denoting the 80{\arcsec} visit splitting distance.  As
  an additional visual cue to guide the user, the ESASky viewer will
  highlight the source currently selected from the Gaia DR3 target
  list.  }
\label{fig:gen_tso_esasky}
\end{figure}

One of the interactive output components of {\tso} is the 
ESASky viewer panel (Figure \ref{fig:gen_tso_esasky}).
ESASky\footnote{\href{https://www.cosmos.esa.int/web/esdc/esasky-help}
{https://www.cosmos.esa.int/web/esdc/esasky-help}} is an online
science driven discovery portal providing full access to the entire
sky as observed with space, ground based, and multi-messenger astronomy
missions \citep{BainesEtal2017paspESASky, GiordanoEtal2018acESASky}.
{\tso} incorporates ESASky primariliy as a viewer panel of the field
of view around the selected science target.

Through the interaction with the Gaia Data Release 3
catalog \citep[DR3][]{GaiaCollab2023aaGaiaDR3Summary}, {\tso} enables
users to search for stellar sources and their properties around a
science target.  This is particularly useful when the science target
is not suitable for acquisition, typically when targets are too bright
for the NIRSpec instrument.

Search queries will deploy a table of stellar targets that are located
within the science target's visit splitting distance
($\sim$80\arcsec).  This table lists the relevant information to
determine the suitability of these sources for target acquisition: target
name, Gaia magnitude, separation from the science target, effective
temperature, $\log(g)$, right ascension, and declination
(Fig.\ \ref{fig:gen_tso_esasky}).  Simultaneously, the ESASky viewer
will display the location of the Gaia targets and the visit splitting
distance area.  Users can then select targets from this list to
perform the target acquisition.

\section{Validation}
\label{sec:validation}

We validated the {\tso} scientific outputs in two ways.  First, we
implemented a test suite using the Python \textsc{pytest} package to
ensure that the calculations that can be directly compared to the
online ETC indeed produce the expected values.  These tests include
the saturation fractions, exposure times, flux rates and signal to
noise ratios, among others.

\begin{figure*}[t]
\centering
\includegraphics[width=\linewidth,clip]{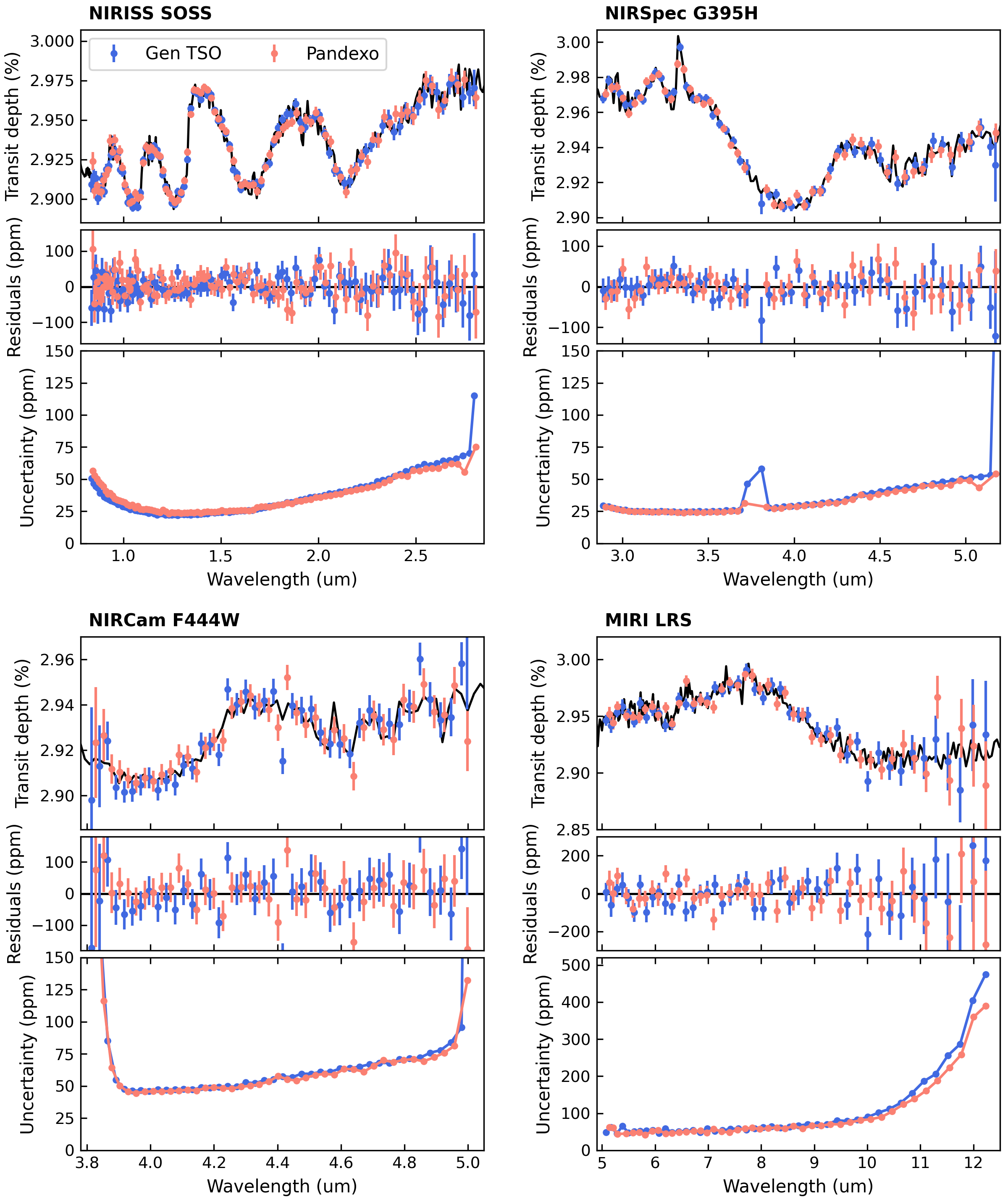}
\caption{
  Benchmarking between {\tso} (blue) and {\pandexo} (salmon) for
  WASP-80b transit simulations with NIRISS (top left panel), NIRSpec
  (top right), NIRCam (bottom left), and MIRI (bottom right).  For clarity, the noisiest data points have been clipped out (when the error > 10 median(error)).
  Table \ref{table:gen_tso_validation} shows the configuration setup.
  For each instrument the top panels show a random realization of the
  simulated spectrum of the true underlying transit model (black
  curve). The simulated spectra have been binned to a resolving power
  of $R=80$ (NIRISS), 90 (NIRSpec), 150 (NIRCam), and 50
  (MIRI). The middle panels show the transit-depth
  residuals compared to a noiseless model.  The bottom panels show
  the corresponding uncertainties of the transit
  depth (note that while the shown transit depths are a
  random stochastic realization, the uncertainties are a deterministic
  calculation).}
\label{fig:gen_tso_validation}
\end{figure*}

More importantly, we also compared the {\tso} depth uncertainty
estimations with those of
{\pandexo} \citep{BatalhaEtal2017paspPandexo}.
{\pandexo} is also an interface to {\pandeia}, designed to compute
signal-to-noise ratios (S/N) for exoplanet time-series observations.
While both tools handle {\JWST} observations, a key difference is that
{\pandexo} also supports time-series simulations for the {\em
Hubble}’s WFC3 instrument.  In contrast, {\tso} is tailored for
{\JWST}, adding features like target acquisition simulations
(Section \ref{sec:acquisition}) and offering easier access to the
targets' system parameters (Section \ref{sec:exo_catalog}) and prior
{\JWST} observations (Section \ref{sec:trexo}).  Also the more
interactive graphical interface of {\tso} facilitates the
visualization and comparison in real time of different stellar SEDs
and planetary spectra.
For our comparison test, we simulated a spectroscopic transit
observation of the warm Jupiter WASP-80b. The host star's SED is a
PHOENIX K5V model ($T_{\rm eff}=4250$~K, $\log g=4.5$), with a Ks
magnitude of 8.351.  The observation has a total duration of 6.0~h
with a transit duration of 2.1~h.
The transit depth spectrum of the planet was computed using the
{\pyratbay} code \citep{CubillosBlecic2021mnrasPyratBay} from a
radiative-thermochemical equilibrium model of a solar-abundance
atmosphere with system parameters consistent with WASP-80b.  This
resulted in a transmission spectrum with a transit depth of $\sim$3\%
and spectral feature amplitudes of $\sim$1 ppt.

We simulated an observation selecting one configuration for each of
the four instruments onboard {\JWST}
(Table \ref{table:gen_tso_validation}).  For each configuration we
selected the fastest readout pattern.  The number of groups was
selected requiring a maximum saturation fraction of 80\%.

{\setlength\tabcolsep{1.0pt}
\begin{table}[h]
\centering
\caption{Instrumental configurations for validation comparison}
\label{table:gen_tso_validation}
\begin{tabular*}{1.0\linewidth} {@{\extracolsep{\fill}} lccccc}
\hline
\hline
Instrument / Mode & Grism & Filter & Subarray & ngroup  \\
\hline
NIRISS / SOSS       & GR700XD & Clear   & SUBSTRIP256 & 3 \\
NIRSpec / BOTS      & G395H   & F290LP  & SUB2048  & 14 \\
NIRCam  / LW Grism  & GRISMR  & F444W   & SUBGRISM64 & 104 \\
MIRI    / LRS       & P750L   & None    & SLITLESSPRISM & 34 \\
\hline
\hline
\end{tabular*}
\end{table}
}

We ran both {\tso} and {\pandexo} using identical input configurations
of the instrument, target, background levels, and spectral extraction
strategy.  Figure \ref{fig:gen_tso_validation} shows the simulated
transit depth spectra and their estimated uncertainties after binning
to a similar resolution.
Both packages produced consistent uncertainties across the spectra,
typically agreeing within $\sim$1--5\%. Small differences are expected
due to variations in how each package processes the data (e.g.,
interpolation between stellar SED and transit depth spectra, rounding
of the number of integration, spectral binning, etc.). The largest
discrepancies seen occur near the edges of the spectra or in the gap between detectors for NIRSpec at $\sim$3.7~{\microns}, which are due
to the difference in binning method from each package.

\section{Summary}
\label{sec:conclusions}

The {\tso} package is a tool designed to compute S/Ns for exoplanet
time-series observations with {\JWST}. While this tool does not
eliminate the need for users to familiarize with the {\JWST}
instrumentation, it significantly simplifies the planning and
simulation of {\JWST} observations by providing an interactive
interface. This interface provides direct access to the database of
confirmed and candidate exoplanets, to the existing catalog of {\JWST}
exoplanet programs, and to the {\pandeia} noise calculator. This
allows users to perform S/N estimations for transit, eclipse, and
target acquisition simulations, while having interactive feedback of
the stellar and planetary spectra, the target's field of view,
integration and saturation times, and all outputs that are typically
provided by the {\JWST} ETC.

We have designed the {\tso} API in a modular manner such that users
have direct access to the astrophysical catalogs and the {\pandeia}
engine functionality through Python scripts of Jupyter notebooks.  For
example, users can search for targets with specific properties or
specific {\JWST} observations.  Users can also perform any of the
{\pandeia} operations shown in this article (e.g., extract stellar
SEDs, calculate exposure and saturation times, or run {\pandeia}).
For practical examples of the API usage, see
Listing \ref{gen_tso_code} and the documentation
at \href{https://pcubillos.github.io/gen_tso/tutorials.html}
{pcubillos.github.io/gen\_tso/tutorials.html}

Future development plans include adding support for
photometric time-series observing modes, facilitating setting custom
parameters for known targets, or enable users to add custom targets not
present in the NASA catalogs.

{\tso} is an open-source package
available for installation via the Python Package Index (PyPI):
\begin{quote}
\indent \texttt{pip install gen\_tso}
\end{quote}
The source code is public and hosted at
Github\footnote{\href{https://github.com/pcubillos/gen\_tso}
  {https://github.com/pcubillos/gen\_tso}}, and the documentation can
be found at \href{https://pcubillos.github.io/gen\_tso}
{pcubillos.github.io/gen\_tso}.

\begin{lstlisting}[float, language=Python, label=gen_tso_code, caption=Sample Python {\tso} script to reproduce the simulation of a WASP-80b transit with NIRspec/G395H shown in Fig.\ \ref{fig:gen_tso_validation}.]
import gen_tso.pandeia_io as jwst
import gen_tso.catalogs as cat
import numpy as np

# Simulate a WASP-80b transit using NASA values
catalog = cat.Catalog()
target = catalog.get_target('WASP-80 b')
sed = jwst.find_closest_sed(
    target.teff,
    target.logg_star,
    sed_type='phoenix'
)

# in-transit and total duration times:
transit_dur = target.transit_dur
t_base = np.max([0.5*transit_dur, 1.0])
obs_dur = 1.75 + transit_dur + 2*t_base


# Planet model spectrum, a plain text file
# containing:  wl(um) and transit depth
depth_model = np.loadtxt(
    'WASP80b_transit.dat', unpack=True,
)

# Initialize the instrument config object
pando = jwst.PandeiaCalculation('nirspec', 'bots')
pando.set_scene(
    sed_type='phoenix', sed_model=sed,
    norm_band='2mass,ks',
    norm_magnitude=target.ks_mag,
)

# Simulate transit TSO remaining at 70% saturation
ngroup = pando.saturation_fraction(fraction=70.0)
obs_type = 'transit'
tso = pando.tso_calculation(
    obs_type, transit_dur, obs_dur, depth_model,
    ngroup=ngroup,
)
\end{lstlisting}

\acknowledgments

{\em We} thank the anonymous referee for his/her time and valuable
comments that improved the manuscript.  PEC acknowledges financial
support by the Austrian Science Fund (FWF) Erwin Schroedinger
Fellowship, program J4595-N.  This research has made use of ESASky,
developed by the ESAC Science Data Centre (ESDC) team and maintained
alongside other ESA science mission's archives at ESA's European Space
Astronomy Centre (ESAC, Madrid, Spain).  This research has made use of
NASA's Astrophysics Data System Bibliographic Services.
The idea for this project was conceived at the ``dotAstronomy 12''
conference in New York, NY, 2023.
PEC thanks contributors to the Python Programming Language and the
free and open-source community, including developers of
{\pandeia} \citep{PontoppidanEtal2016spiePandeia},
{\pandexo} \citep{BatalhaEtal2017paspPandexo},
\textsc{Shiny}, \textsc{plotly}, 
\textsc{Numpy} \citep{HarrisEtal2020natNumpy},
AASTeX6.2 \citep{AASteamHendrickson2018aastex62},
and
\textsc{bibmanager}\footnote{
\href{https://bibmanager.readthedocs.io}
     {https://bibmanager.readthedocs.io}}
\citep{Cubillos2020zndoBibmanager}.

\bibliography{gen_tso}

\end{document}